\documentclass[3p,times]{elsarticle}

\usepackage{ecrc}

\volume{00}

\firstpage{1}

\journalname{Procedia Computer Science}

\runauth{}

\jid{procs}

\jnltitlelogo{Procedia Computer Science}

\usepackage{amssymb}

\usepackage[figuresright]{rotating}
\usepackage{amsmath}
\usepackage{amssymb}
\usepackage{tabularx}
\usepackage{subfloat}
\usepackage{subfig}
\usepackage{multirow}
\begin{document}

\begin{frontmatter}

\dochead{}

\title{A Survey on E-Commerce Learning to Rank}

\author[First]{Md. Ahsanul Kabir}
\ead{mdkabir@iu.edu}

\author[Second]{Mohammad Al Hasan}
\ead{alhasan@iupui.edu}

\author[Third]{Aritra Mandal}
\ead{arimandal@ebay.com}

\author[Fourth]{Daniel Tunkelang}
\ead{dtunkelang@ebay.com}

\author[Fifth]{Zhe Wu}
\ead{zwu1@ebay.com}

\begin{abstract}
In e-commerce, ranking the search results based on users' preference is the most 
important task. Commercial e-commerce platforms, such as, Amazon, Alibaba, eBay,
Walmart, etc. perform extensive and relentless research to perfect their search 
result ranking algorithms because the quality of ranking drives a user's decision to purchase or not to purchase an item, 
directly affecting the profitability of the e-commerce platform. In such a commercial
platforms, for optimizing search result ranking numerous features are considered, which
emerge from relevance, personalization, seller's reputation and paid promotion.
To maintain their competitive advantage in the market, the platforms do no publish their
core ranking algorithms, so it is difficult to know which of the algorithms or which
of the features is the most effective for finding the most optimal search result
ranking in e-commerce. No extensive surveys of ranking to rank in the e-commerce domain
is also not yet published. In this work, we survey the existing e-commerce learning to
rank algorithms. Besides, we also compare these algorithms based on query relevance
criterion on a large real-life e-commerce dataset and provide a quantitative analysis.
To the best of our knowledge this is the first such survey which include an experimental
comparison among various learning to rank algorithms.
\end{abstract}

\end{frontmatter}

\section{Introduction}

Ranking is a central problem in information retrieval. Ranking result typically depends on the relevance between documents and queries, that's why in some references ranking is also called relevance ranking. To illustrate, given a query, the document will have higher ranking if it is more relevant to the query than other documents. If there is a large number of documents associated with a query, the top ranked documents are the most important documents as well. 

While learning to rank (LTR) has been the focus of information retrieval for maximizing search relevance~\cite{KeywordSearch2013,Diversifying2009, ProductSearch2011,facet2013,Wilcoxon1945IndividualCB,EnhancingProductSearch2012}, the applications of LTR include collaborative filter, document retrieval, definition finding, key term extraction, important email routing, product rating, sentiment analysis, and anti web spam. However, in the e-commerce domain, the purpose of LTR is to help the buyers to find and purchase their intended products. The buyers' intent advocates the demand side while the listings by the e-commerce site represent the supply side of the platform. The goal of the e-commerce LTR is to provide the most relevant products for the supply and the demand side and to maximize the chance to generate revenue. While both a web search engine and an e-commerce site intend to ensure user satisfaction by presenting the most relevant products that a user is searching for, an e-commerce site maximizes revenue or sales, and minimizes the inventory cost showing the most relevant products. Additionally, the feature selection strategy of the two differs because the e-commerce queries are often shorter, lack syntactic structure and product property dominant than the web queries. Moreover, the ranking of products for the e-commerce queries are performed for user signals, like click rate, add-to-cart rate, and purchased rate.

Due to the importance of e-commerce product ranking, quite a bit of research is performed in this area~\cite{ltr_devapujula,ltr_2021_ali,ltr_2021_expedia,ltr_2018_ebay}. However, due to business privacy, and competition among the companies, the methods of LTR remain private. For the same reason, and user privacy most of the datasets are not public in this domain. There is a survey illustrating the challenges, business insights,  and research opportunity in e-commerce domain~\cite{survey_search_recom} for relevance learning, product classification, taxonomy classification, recommendations, search personalization, learning to rank etc. However, the goal of this paper is to provide a general research idea in e-commerce domain. There is another survey~\cite{iqbal2019production} which focuses on data processing, representation learning, candidate selection, and general idea of product ranking. But illustrating the models details, experimenting with the models, and ranking among the models are not emphasized much.
There is hardly any research to illustrate and rank the current LTR methods based on experiment in e-commerce domain.
So the contributions of this paper are the following --

\begin{itemize}
    \item To our best knowledge we are the first to illustrate the LTR methods in details describing the e-commerce specific challenges and probable solution.
    \item We perform experiments with the LTR methods and rank them based on the performance in a real e-commerce dataset.
\end{itemize}

We organize the paper as follows -- In Section 2 we formulate the problem of LTR using a common framework, then we divide the LTR methods into three groups and illustrate the mechanism of each method demonstrating the loss function each method is solving. Next we illustrate the e-commerce specific challenges in Section 3, and the evaluation metrics used in LTR methods in Section 4. After that in Section 5 we demonstrate the possibility of finding available datasets in e-commerce domain, and finally in Section 6 we provide the experiment results for all the illustrated methods and rank them based on the performance in an e-commerce dataset.

\section{LTR Framework} \label{exp}
To demonstrate the machine learning frameworks, we first formally define the LTR problem in e-commerce domain. Let a training dataset contain $n$ training queries $\{q_i\}_{i=1}^{n}$. The corresponding products for $q_i$ be \{$p^{(i)}_{j}$\} $_{j=1}^{m^{(i)}}$ such that $p^{i}_{1} \succeq p^{i}_{2} ....\succeq p^{i}_{m^{(i)}}$ in terms of relevance and $m^{(i)}$ be number of relevant product for $q_i$. The goal is to design a ML framework which learns to rank from the training instances and perform best in the test data. Each framework approaches the LTR problem in two ways. First, they define different input and output spaces. Input spaces are contingent on feature selection and representation learning of query and products. The output space is to facilitate the learning process based on ranking the documents with respect to query document relevance. Second, each framework has a particular heuristic to learn a scoring function which can rank the products with respect to the corresponding e-commerce query. To learn the scoring function each method minimizes a distinguished loss function $\mathcal{L}$. 
Note that, there are quite a few research available for LTR in information retrieval~\cite{Swezey2020PiRankLT, Pang2020, AI2019}. However, these methods are not e-commerce specific. On the other hand, there are other methods available in e-commerce domain which are designed to meet specific requirements~\cite{Pfannschmidt2018DeepAF,Pei2019}. In this research we want to demonstrate the important existing methods along with the loss functions each method solves. Our main goal is to provide a general idea of LTR methods, and in the next subsections we illustrate them.

\begin{figure*}
    \centering
    \includegraphics[width=0.8\linewidth]{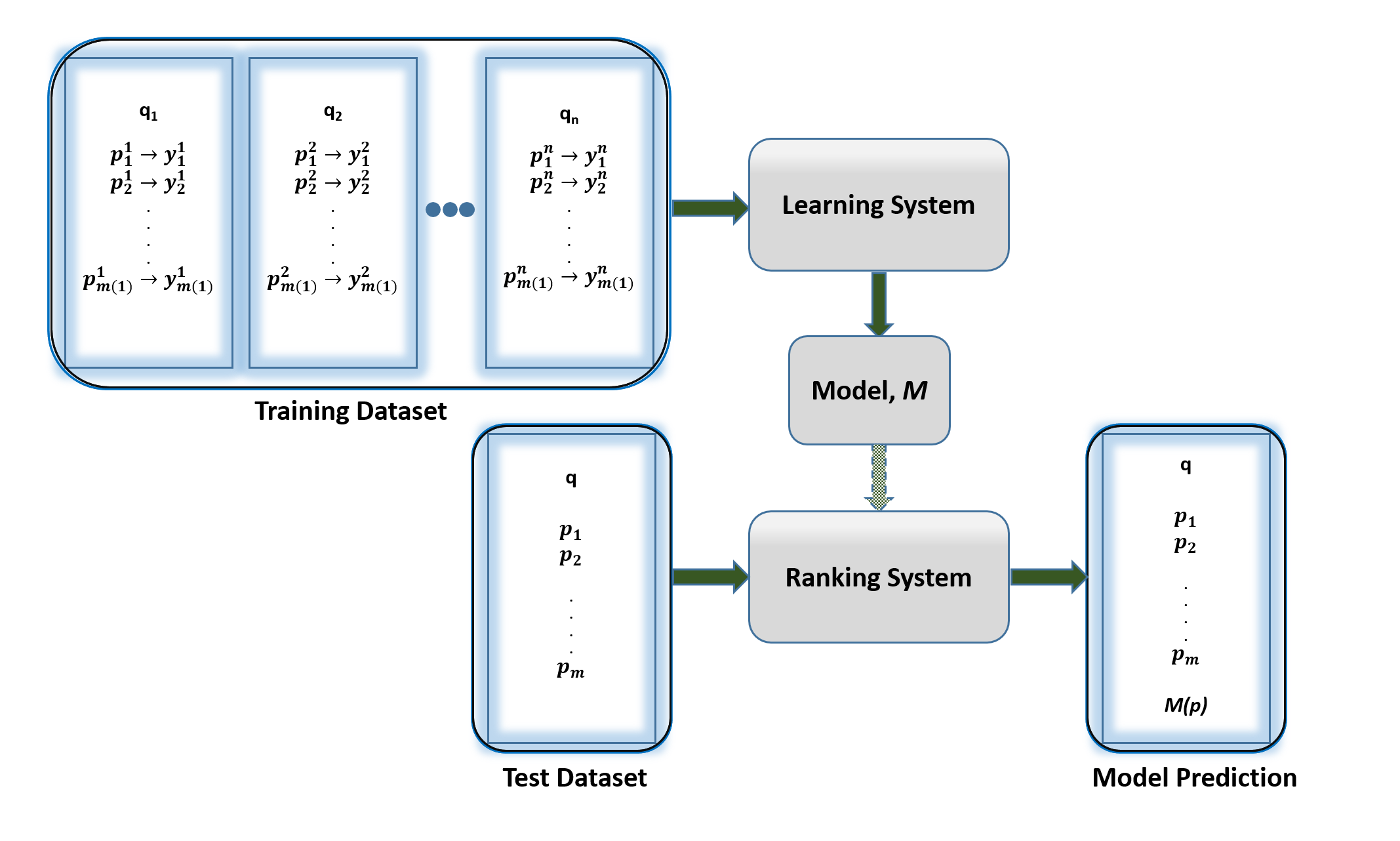}
    \caption{Illustration of a LTR model to perform e-commerce product ranking task.}
    \label{fig:model}
\end{figure*}

\subsection{Pointwise Approaches}
Pointwise approaches \cite{ARSM2019,svm_ordinal_regression_2005,BroadNarrow2019,Gaussian_processes_2005,logistic_regression_1992,PolynomialRegression1989,Large_Margin_2002, ltr_2018_ebay} assign each query, product pair a relevance score. The ranking is performed in a straightforward way checking whether existing
learning methods can be directly applied.
To illustrate, a $d$ dimensional vector, $x^{(i)}_{j}$ is constructed for each query product pair ($q_i$, $p^{(i)}_{j}$) as input space. In addition, a list of floating point value scores, $y^{(i)}$ where $y^{(i)}=[y^{(i)}_{1},..., y^{(i)}_{m^{(i)}}]$ is produced for each product as output space. The score $y^{(i)}_{j}$ represents the relevance degree of product $p^{(i)}_{j}$ to query $q_i$. $y^{(i)}_{j}$ is calculated generally from buyer interaction of products in a e-commerce site. Once $\{(x^{(i)}, y^{(i)})\}_{i=1}^{n}$ is constructed from training dataset, the goal is to learn the mapping between $x^{(i)}$ and $y^{(i)}$. 
Among the regression methods linear regression~\cite{Free:2005}, polynomial regression~\cite{poly2012} methods reduce squared loss which is defined by the next equation.
\[\mathcal{L} = \sum_{j=1}^{m^{(i)}}{(y^{(i)}_{j} - f(x^{(i)}_j))^2}\]
Note that for linear regression $f(x^{(i)}_j)) = w^T.x^{(i)}_j + b$ where $w$
is a trainable weight vector of dimension $d$, and $b$ is the bias. In contrast, for polynomial regression, $f(x^{(i)}_j)$ is defined in the following equation where $w_k$ is the $k$'th value of a $d$ dimensional weight vector $w$, and similarly $x^{(i)}_{jk}$ is the $k$'th value of the $d$ dimensional vector $x^{(i)}_{j}$.
\[ f(x^{(i)}_j) = \sum_{k=1}^{d}(w_k.(x^{(i)}_{jk})^j) + b\]

While regression methods work with continuous value for $y^{(i)}_{j}$, there are some researches~\cite{li2007} which approach the ranking problem as a classification problem converting $y^{(i)}_{j}$ to either a binary variable indicating whether a product is relevant to a query, or ordinal multi-class variable for most relevant to least relevant classes. Mathematically, for binary output space $y^{(i)}_{j}$ is converted to $c^{(i)}_{j}$ such that $c^{(i)}_{j}$ = 1 if $p^{(i)}_{j}$ is relevant to $q_i$ otherwise $c^{(i)}_{j}$ = 0. On the contrary, for multi-categorical ordinal output space $c^{(i)}_{j}$ = $\{l_c\}_{c=1}^{K}$ where $l_c$ is a category and $K$ is the total number of categories. Furthermore, $l_K$ $\succ$ $l_{K-1}$ .... $\succ$ $l_1$ in terms of relevance.
Both the classification and ordinal regression problem can be solved by support vector machine (SVM)~\cite{svm1995}. The goal is to adopt a linear scoring function. The constraints actually require every
product to be correctly classified into its target ordered category.
For instance, for a product of category $l_c$, $w^T.x^{(i)}_{j}$ is expected to exceed threshold $b_{c-1}$ but get smaller than $b_c$, with certain soft margins $1 - \epsilon^{(i)*}_{j,c-1}$, and $1 - \epsilon^{(i)}_{j,c}$ respectively. The optimization function can be illustrated as the following way.

\begin{equation}
\begin{aligned}
\min \quad & ||w||^2+C\sum_{i=1}^{n}\sum_{j=1}^{m^{(i)}}\sum_{c=1}^{K-1}(\epsilon^{(i)}_{j,c}+\epsilon^{(i)*}_{j,c+1}) \\
\textrm{s.t.} \quad & w^T.x^{(i)}_{j}-b_c\leq-1+\epsilon^{(i)}_{j,c} \quad & \textrm{if} \quad & c^{(i)}_{j}=l_c\\
\quad & w^T.x^{(i)}_{j}-b_c\geq 1-\epsilon^{(i)*}_{j,c+1} \quad & \textrm{if} \quad & c^{(i)}_{j}=l_{c+1}\\
  &\epsilon^{(i)*}_{j,c+1}\geq0,\textrm{ }\epsilon^{(i)}_{j,c}\geq0   
\end{aligned}
\end{equation}

However, there are more advanced methods available in the literature. For intstance, Gradient Boosted Methods~\cite{FRIEDMAN2002367,Palotti2016LearningTR}, and Random Forest~\cite{FRIEDMAN2002367}  work on ensemble learning with respect to multiple weak learners. The basic idea is, if there are multiple shallow learners available, and the outputs are observed carefully, the probability of making mistake gets reduced. While both Gradient Boosting, and Random Forest are ensemble methods, they differ on gathering the weak learners' outputs on final decision. Random Forest
ensembles the outputs of the weak learners using bagging. On the contrary, Gradient boosting ensembles based on gradient boosting.  

\subsection{Pairwise Approaches} 
Like the point-wise approaches, the pairwise approaches~\cite{ranknet2005,lambdarank2006,hu2019,frank2007,dai2011,Cerrato2020,jia2021} do not work with relevance scores of all the titles with respect to the associated query. In fact the rank of products in e-commerce domain are more important than relevance scores. That's why instead of relevance scores, pair of titles are chosen from all the $m^{(i)}$ titles, and models are trained based on relative ranking of the titles. Due to the emphasis of relative ranking, pairwise approaches are closer to the concept
of ranking than the pointwise approaches. 

Among the pairwise approaches, probably the most famous one is RankNet~\cite{ranknet2005}. RankNet solves a simple probabilistic cost function and can be implemented combining neural network with stochastic gradient descent method. To illustrate RankNet, recall the query $q_i$, and the two associated products $p^{(i)}_{j}$, $p^{(i)}_{k}$ such that $p^{i}_{j} \succ p^{i}_{k}$. In order to compare the relative importance of the two products with respect to $q_i$, a rank function $f: \mathcal{R}^{D} \rightarrow \mathcal{R}$ is designed such that $f(p^{i}_{j})$ > $f(p^{i}_{k})$. In other words, $f$ converts a $D$ dimensional feature vector extracted from product and query to a real-valued scalar which preserves the relative ranking of the two products. Let $z$ be $f(p^{i}_{j})$ - $f(p^{i}_{k})$, then $z$ > 0 to ensure $p^{i}_{j} \succ p^{i}_{k}$. Next the probability, $P_{jk}$ of 
$p^{i}_{j} \succ p^{i}_{k}$ is calculated using $z$, and sigmoid activation function with scaling factor $\sigma$.

\[P_{jk} = \frac{1}{1+e^{- \sigma z}}\]

Like any binary classification with sigmoid activation function, to rank $p^{i}_{j}$ higher than $p^{i}_{k}$, $P_{jk}$ needs to be greater than 0.5. Presenting the probabilistic framework improves the consistency of the model, and a cross entropy (CE) loss can be designed for optimization problem.

\[C_{jk} = - \hat{P}_{jk} \log P_{jk} - (1 - \hat{P}_{jk}) \log (1 -P_{jk})\]

The loss function penalizes if the target order label $\hat{P}_{jk}$ and the predicted output $P_{jk}$ does not match. To demonstrate $\hat{P}_{jk}$ for a given query, let $S_{jk}$ $\epsilon$ \{1,0,1\} be defined to be 1 if $p^{i}_{j}$ is more relevant than $p^{i}_{k}$,
-1 if $p^{i}_{j}$ is less relevant than  $p^{i}_{k}$, and 0 if both the products have the same label. $\hat{P}_{jk} = \frac{1}{2} (1 + S_{jk})$. By using these definitions CE loss function can be illustrated as the following equation.

\[C_{jk} = \frac{1}{2} (1 - S_{jk}) \sigma z + log (1 + e^{-\sigma z}) \]
A neural network is designed to reduce the CE loss function using gradient descent algorithm. 

There are several improvement methods of RankNet, among which LambdaRANK~\cite{lambdarank2006} from Microsoft research deals with the weakness of the cost function. Note that the cost function RankNet solves is based on penalizing count of flipped of product orders for a given query. However, although the loss function is convex, and differentiable minimizing the loss function does not always improves NDCG score. To recovery from that, LambdaRANK is proposed introducing a loss function which is rewarded based on improvement of NDCG. To illustrate, consider the CE loss function of RankNet, and let $s_j = f(p^{i}_{j})$, and $s_k = f(p^{i}_{k})$. \[\lambda_{jk} = \frac{\delta C_{jk}}{\delta s_j} = \sigma (\frac{1}{2} (1 - S_{jk}) - \frac{1}{1+e^{-\sigma z}}) \] 
For RankNet these lambda values are used to speed up the gradient descent calculation. But for LambdaRANK, the equation is slightly modified as the following.

\[\lambda_{jk} = \frac{\delta C_{jk}}{\delta s_j} =  \frac{-\sigma}{1 + e ^ {\sigma z}} |\Delta_{NDCG}| \]
That means, the lamdas are penalized by how much swapping a pair of products improve NDCG while keeping the other pair orders unchanged. In stead of NDCG, other evaluation metrics can also be used in the loss function.

Another improvement of RankNet is FRank~\cite{frank2007}. Note that the cross-entropy loss of RankNet has a non-zero minimum which leads to have some loss irrespective of any model. FRank solves this problem proposing a new loss function, {\it fidelity loss}. 

\[C_{jk} = 1 - \sqrt{\hat{P}_{jk} P_{jk}} - \sqrt{(1 - \hat{P}_{jk})  (1 -P_{jk})}\]
The idea was borrowed from quantum mechanics to measure
the difference between two probabilistic states of the target probability and the modeled probability.

One of the widely used improvements of RankNet is LambdaMART~\cite{lambdarank2006}. The idea is to introduce gradient boosting with LambdaRANK. Each weak learner of decision tree solves the loss function of LambdaRANK. The final decision is made combing the outputs of the weak learners using boosting. The combined loss function is constructed based on outputs of all the weak learners. The final loss function is optimized using gradient descent algorithm. Additionally, LambdaMART is further improved in another research~\cite{dai2011} which they refer as Augmented Lagrangian (AL). The main idea is to improve the performance of the weak learners. They use surrogated cost functions to optimize the evaluation metrics like NDCG. Suppose, in the case of NDCG the cost of LambdaMART is used and additionally an upper bound is set to restrict the NDCG based cost. For the given upper-bound (UB) $b$, the cost function is constrained such that $C^t$ <= $b^t$ where $t \in [1,T]$.  This makes it a constrained optimization problem, and this loss function of AL improves the loss function of LambdaMART by a large margin. However, in this research we could not compare with AL as the code is not publicly available.

In the literature, there exists SVM based model as well for pairwize ranking model. Recall the query $q_i$, the two associated products $p^{(i)}_{j}$, $p^{(i)}_{k}$, and the corresponding label $y^{(i)}_{jk}$. If $p^{(i)}_{j}  \succ p^{(i)}_{k}$, $y^{(i)}_{jk}$ = 1, otherwise $y^{(i)}_{jk}$ = -1.
A soft margin SVM model can solve this problem minimizing hinge loss.
The loss function and constraints are described by the next terms.


\begin{equation}
\begin{aligned}
\min \quad & ||w||^2+C\sum_{i=1}^{n}\sum_{j,k,y^{(i)}_{jk} = 1} \epsilon^{(i)}_{jk} \\
\textrm{s.t.} \quad & w^T.(x^{(i)}_{j} - x^{(i)}_{k})\geq 1 - \epsilon^{(i)}_{jk} \quad & \textrm{if} \quad & y^{(i)}_{jk} = 1\\
  &\epsilon^{(i)}_{jk}\geq0  
\end{aligned}
\end{equation}

The difference between Ranking SVM and SVM lies in
the constraints, which are constructed from product pairs. If $y^{(i)}_{jk}$ = 1, and $w^T.x^{(i)}_{j} \ge w^T.x^{(i)}_{k}$ by margin of 1, then there is no loss, otherwise the loss is $\epsilon^{(i)}_{jk}$

\subsection{Listwise Approaches} \label{list}
For listwise approaches~\cite{burges2010from,xia2008,listnet2007,plistmle2014,taylor2008softrank,Xu2007AdaRankAB,Lan2009,Sharma2020,Shi2010,Lan2009,Li2021}, two types of sub-categories exist in the literature. For the first sub-category, it is only known whether a title is relevant to a query. While the titles are not ranked among them, no relevant scores are provided for a query, title pair. On the contrary, for the second sub-category, a permutation of a list of titles are available where the titles are arranged in terms of relevance order. In e-commerce domain, the ranking among titles is important. It is less likely for a customer to buy the products which are far apart from the viewed page, although the titles are relevant. That's why we illustrate the methods in the second sub-category. The listwise approaches are the closest in terms of ranking because the loss functions can distinguish which ranking is better than other. This is why the listwise approaches are the most promising methods for LTR problems.

The first listwise method we describe in this paper is ListNet~\cite{listnet2007}.
To illustrate, recall the query $q_i$, and the ${m^{(i)}}$ number of products. Let a permutation of the products be \{$\pi^{(i)}_{j}$\} $_{j=1}^{m^{(i)}}$ associated with $q_i$, and the actual index of the products in the ground truth rank order be $l^{(i)}_y$ = \{${o}^{(i)}_{j}$\} $_{j=1}^{m^{(i)}}$.
 Let a rank function $f: \mathcal{R}^{D} \rightarrow \mathcal{R}$, such that $f(x^{(i)}_{j})$ = $s^{(i)}_{j}$ where $x^{(i)}_{j}$ is a representation vector of product $\pi^{(i)}_{j}$. A probability function is defined to calculate the probability of a rank order. For instance, the probability of the rank order \{$p^{(i)}_{j}$\} $_{j=1}^{m^{(i)}}$ is defined by the next equation.

\[P_s = \prod_{k=1}^{m^{(i)}} \frac{\psi(s^{(i)}_{k})}{\sum_{u=k}^{u=m^{(i)}}\psi(s^{(i)}_{u})}\]

Here $\psi$ is a monotonically increasing function which can be linear,
exponential, or sigmoid. While each item  $\frac{\psi(s^{(i)}_{k})}{\sum_{u=k}^{u=m^{(i)}}\psi(s^{(i)}_{k})}$ in the equation is a conditional probability value,  $P_s$ gets the largest probability value when the ranking order is appropriate. In the same fashion, another probability value, $P_y$ is calculated for the actual ranking of the products with respect to $q_i$. Based on these two probability functions, a K-L divergence loss function can be derived for ListNet which the model optimizes. 

While ListNet is a very closer model for actual ranking, due to K-L divergence loss, the train complexity is exponential to $m^{(i)}$.
So to avoid the complexity, if number of elements in the permutation are kept smaller, information about the permutation
is significantly lost and the effectiveness of the ListNet algorithm
is questionable. Another approach ListMLE~\cite{xia2008} is proposed to avoid the drawbacks of ListNet. For each
query $q_i$, for the products in the ground truth permutation, a probability distribution based on the output of the scoring function is defined where $x{_l}^{(i)}_{k}$ is a representation vector of a product in the k'th position of actual ranking.

\[P_{y|x} = \prod_{k=1}^{m^{(i)}} \frac{\psi(f(x{_l}^{(i)}_{k}))}{\sum_{u=k}^{u=m^{(i)}}\psi(f(x{_l}^{(i)}_{u}))}\]

For ListMLE in stead of K-L divergence loss, it uses the negative log likelihood of $P_{y|x}$. As both K-L divergence loss, and negative log likelihood loss are convex, and differentiable, both model can be easily optimized. But ListMLE converges faster than ListNet as the train complexity is $O(m^{(i)})$.

Although both ListMLE, and ListNet can solve LTR challenges, position of the products are not emphasized in any of these models. For instance, if there is a flip of order of two products in early position of the permutation, the cost should be higher, as it is a matter of purchase. Note that, the latter products are not as important as the former ones, so the cost of error of both the cases should not be the same. That's why another model position aware ListMLE~\cite{plistmle2014} (p-ListMLE) is proposed. 
The loss function for p-ListMLE is defined by the following equation.

\[L_p = \sum_{k=1}^{m^{(i)}} \alpha(i)(- f(x{_l}^{(i)}_{k})) + \log (\sum_{u=k}^{m^{(i)}} \psi (f(x{_l}^{(i)}_{u}) ))\]

where $\alpha(.)$ is a decreasing function, i.e. $\alpha(i) > \alpha(i + 1)$. $L_p$ introduces more penalty in the earlier positions of the actual permutation.

Among the advanced methods, there exists a reinforcement based method DeepQRank~\cite{Sharma_DeeqRank} which solves the LTR problem by a 
listwise Markov Decision Model. The loss function is calculated from how much gain is achieved when a product is added from unordered to ordered list. Recall the query $q_i$, and let there be a ranked list for predicted products, $R_L$ initially empty and an unordered list $U_L$ initially containing all the associated products of $q_i$ in arbitrary order. A state for Markov model is consist of $R_L$, and $U_L$ and the state is changed when a product, $p^{(i)}_{j}$ from $U_L$ is moved to $R_L$. When the product is moved to change the state, a timestamp $t_i$ initially 0 is increased. Let the relevance score of $p^{(i)}_{j}$ be $rel^{(i)}_{j}$ such that $rel^{(i)}_{j}$ $\epsilon$ \{$1, 2, ...,m^{(i)}$\}. The rank score of highly relevant products are higher. A reward function is then designed based on Discounted Cumulative Gain which is defined next.

\[r_i = \frac{rel^{(i)}_{j}}{t_i}\]

The policy of the Markov model is to transfer products from $U_L$ to $R_L$ until $U_L$ become empty maximizing $r_i$. The model returns $R_L$ as output products for $q_i$.

\begin{table}
\small
\renewcommand{\arraystretch}{1.3}
 \centering\caption{LTR methods Group}

    \begin{tabular}{c|  c}
    \bf Group & \bf Description \\\hline
    Pointwise & Each query, product instance is a point, and relevance score or binary relevance label is the target score/label. \\ 
    Pairwise &  Given a query and two product items, the goal is to predict whether the first item is more relevant to the query.\\
     Listwise & Given a query and a permutation of products, the goal is to design a method which learns the relative ordering.\\
    \hline
    \end{tabular}
 
\label{table:groups}
\end{table}

In short, each group of methods solve the LTR problem in different ways. For pointwise methods, the relative ranking among product items is not learned directly. While the pairwise methods aim to learn the ranking system based on the relevance difference between two items with respect to the query, the listwise approaches learn from the permutation of the products in a list based on ordering. Table~\ref{table:groups} illustrates the differences in a nutshell. In the next section we illustrate the e-commerce specific challenges and the ways to recover.

\section{E-Commerce Specific Challenges} The methods so far we described can be applied to any LTR methods. However, e-commerce LTR methods need to solve extra challenges which are not the same as in other domains. 
First, in Web domain if anyone search with a query, normally the targeted web-item appear among the top 3 results. At least 90\% of time the search results are found in the first page in Google. It is said that the best hiding place for a dead body is really on the second page of Google or beyond. For example, if a person searches for \textit{"IUPUI"}, the website of the university will be found in the first result, so it is not even necessary to check the next items in the search result. On the contrary, in e-commerce domain people like to browse more before buying a product as it is a monetary matter. The relevant products can even be in the second and third pages. The average number of products to browse in eBay is top 20. This makes a major difference between e-commerce and web LTR. To solve this specific issue, we need to beer in mind that in Web domain the top results are the most important ones so the mistakes in later documents penalize less than those of top results. But in e-commerce domain this is not the case as user spends time while browsing items. The weights of the error needs to adjust for that. For instance, ListMLE is more suited to e-commerce domain than p-ListMLE as the later products in the search space are also important.

\begin{figure}
    \centering
    \includegraphics[width=0.7\linewidth]{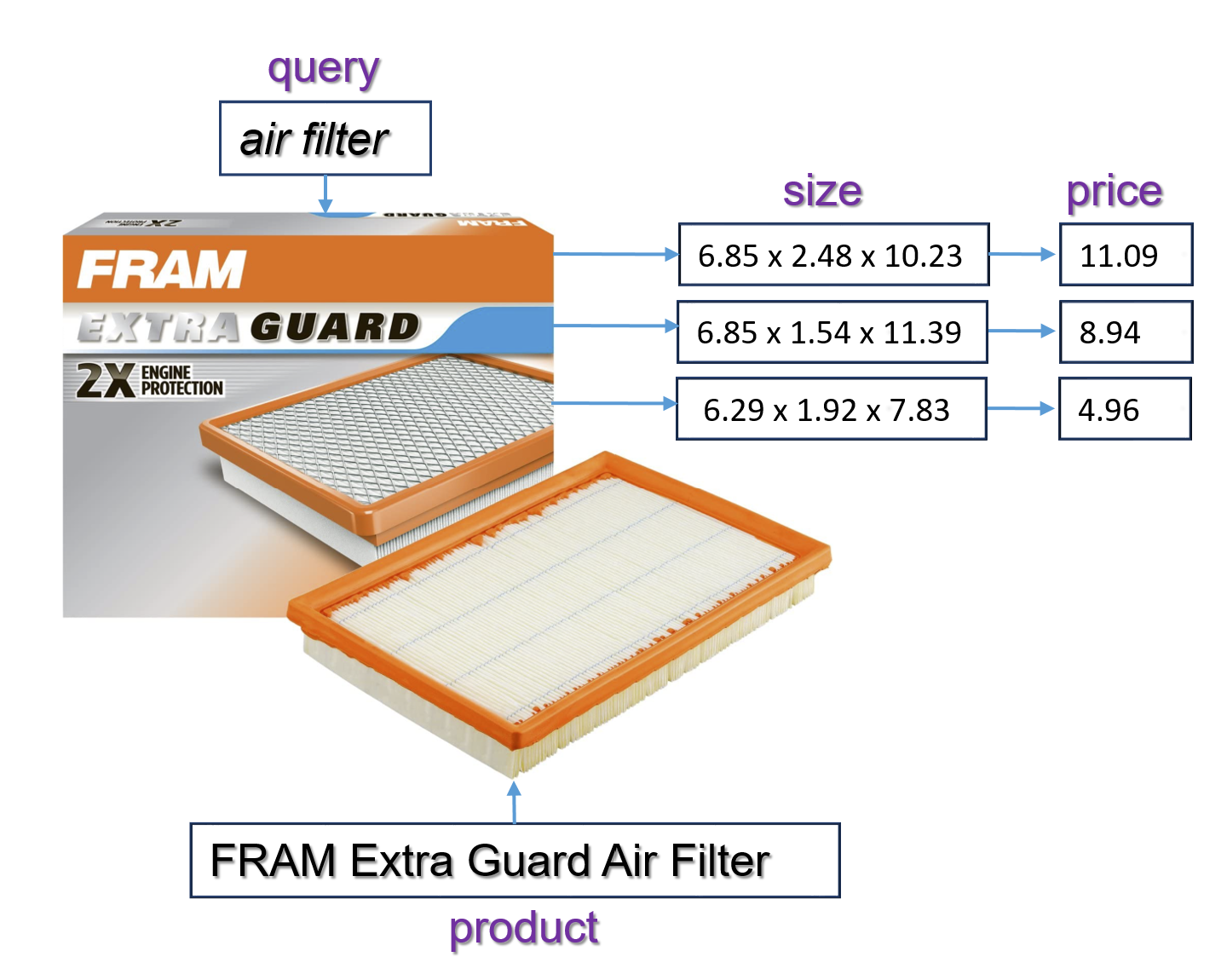}
    \caption{Illustration of a same product appearing multiple times of a single query.}
    \label{fig:sim-query}
\end{figure}

Second, in web domain the search results are distinct, which can provide different relevant scores for every result. This distinguishability can help a proper ranking of the results. On the contrary, products in the e-commerce domain do not always follow this criteria. Same product can be sold by different sellers. Moreover, depending on the features of a product, different users buy different varieties of the same product. To illustrate with an example, consider the diagram in~\ref{fig:sim-query}. If a buyer searches with the query \textit{"air filter"}, quite a few products are shown up in Amazon. Among these \textit{"FRAM Extra Guard Air Filter"} appears three times if the first two pages are explored. Note that, although the product is the same, the size is not the same among these three products. While the size changes, the price values change rapidly too. The e-commerce sites need to overcome this challenge. On the other hand, there is exactly one website for \textit{"IUPUI"}. While the relevance score of the product \textit{"FRAM Extra Guard Air Filter"} are exactly similar for the three cases, the \textit{"Computer Science Department"} of \textit{"IUPUI"} does not have exact same relevance score to other departments in web based search. To overcome this challenge, dataset construction and feature selection needs to be carefully designed. As the feature of the same product changes the relevance score, this must be taken care of when the dataset is designed.

Third, e-commerce product ranking is not static, rather it changes rapidly. The ranking changes due to change of demand, user choice, trend, improvement of the other alternative products, change of prices etc. But web result ranking do not change as rapidly as e-commerce. For instance, the query \textit{"IUPUI"} always require to find the website of \textit{"IUPUI"} at the top. But \textit{"air filter"} finds different types of air filters over time. In other word there can be some different filters other than \textit{"FRAM Extra Guard Air Filter"} may come to the top position depending on the changed status of market. This is an issue which is very important when a model is applied to production. Note that, as gigabytes of the new data are added every day, the model need to be scalable and fast enough with the changed snapshot of ranking data.

Fourth, e-commerce queries are shorter and products property dominant. The syntactic structure of the queries are not as important as web queries. Often web queries are more like natural language sentences. Product band, feature, symbol etc are some of the properties of e-commerce queries. User's previous search history, choice, session are other driving forces for e-commerce ranking. Different users can purchase different products searching with exact same query. The understanding of two different buyers can vary based on the culture, environment, choice, and customs of individuals. Moreover, it is often found that exact matching of a title of the product with query does not necessarily require the product to be in the top in the search. So feature selection of e-commerce queries are completely different than web queries. Even the representation of the queries itself is a challenging task. The existing embedding methods~\cite{bert:Embedding,fasttext_Embedding} trained in the conventional Google and Wikipedia corpus do not capture the query demands. To represent query there is an existing research~\cite{ordsim_mdkabir} which claim that spherical embedding trained for queries, and titles along with Wikipedia and Google corpus can capture query-query similarity better.  We solve this specific challenge retraining the embedding methods in e-commerce specific domain.

\section{Evaluation Metrics} 
Evaluation metrics in the e-commerce domain are normally Normalized Discounted Cumulative Gain (NDCG), Mean Average Precision (MAP), Expected Reciprocal Rank (ERR), Mean reciprocal rank (MRR), Spearman's Rho ($\rho$) etc. 

\subsection{NDCG}  Normalized Discounted Cumulative Gain (NDCG) is used to evaluate any LTR system. NDCG is actually a ratio between two measures i.e., Discounted Cumulative Gain (DCG), and Ideal Discounted Cumulative Gain (IDCG). The purpose of the DCG is to penalize the highly relevant products coming later in search result. Recall the query $q_i$, and the relevant scores of the predicted products $rel^{(i)}_{j}$. The DCG score is calculated by the next equation.

\[DCG = \sum_{j=1}^{m^{(i)}} \frac{rel^{(i)}_{j}}{log_2 (j+1)}\]

The Ideal DCG is then calculated for top $m^{(i)}$ products according to decreasing order of relevant scores in sorted order. To illustrate with an example let $m^{(i)} = 5$, the predicted products are $p^{i}_1$, $p^{i}_2$, $p^{i}_3$, $p^{i}_4$, and $p^{i}_5$ with actual relevant scores 0.6, 0.4, 0.5, 0.3, and 0.4 respectively. Then DCG = $\sum_{j=1}^{j=5} \frac{rel^{(i)}_{j}}{log_2 (j+1)}$ = 1.386. To calculate Ideal DCG let there be another two products with relevance 0.5, and 0.4. All of these seven products are needed to be sorted in descending order, and then top 5 relevant scores are chosen for DCG calculation. The best scores are 0.6, 0.5, 0.5, 0.4, 0.4 and Ideal DCG scores are calculated based on these
scores. The IDCG score = 1.492, and the NDCG score = DCG / IDCG = 0.9288.

Note that, the NDCG score for this particular instance is very high, as the similarity scores have low variance. In this work, we use NDCG slightly modifying the original idea so that it fits with e-commerce domain. First we work with the original relevant scores. Note that the scores are 0.6, 0.5, 0.5, 0.4, 0.4, 0.3 in sorted order. We convert the values to integer rank such that the highest value gets the highest rank, and lowest value gets lowest rank 1. So the rank values for the similarity sores are 4, 3, 3, 2, 2, 1.
Now, to calculate DCG we use this rank values instead to uplift the variance. Finally, we use another version of DCG calculation to penalise the mistakes more than before.

\[DCG = \sum_{j=1}^{m^{(i)}} \frac{2^{rank^{(i)}_{j}}}{log_2 (j+1)}\]

The new DCG is calculated for rank values 4, 2, 3, 1, and 2 which are the rank scores for the similarity values 0.6, 0.4, 0.5, 0.3, and 0.4 respectively.
DCG = 24.932, IDCG = 28.317, and NDCG = 0.88 which is lower than 0.9288. However, if the similarity values are distinct, the NDCG score will be much less than the obtained value as for the exact same similar score we introduce constant rank value. In our case the similarity values are always distinct due to the floating point nature, and we use the modified version of NDCG calculation for better evaluation of e-commerce ranking.

\subsection{MAP} While Mean Average Precision (MAP) is used largely for object detection methods in computer vision, MAP can be used to evaluate the performance of LTR algorithms as well. Note that NDCG can not properly describe the performance of a LTR model when the standard deviation of relevance scores are very small. In that sense, MAP is a good candidate for evaluation metrics. To illustrate, MAP is contingent upon precision which demonstrates what ratio of the predicted products are relevant. Note that relevant scores are not used directly for calculating precision. Additionally, precision at $K$ ($prec@K$) signifies what ratio of first $K$ predicted products with respect to a query are relevant. Finally, average precision is calculated varying $K$ for all the predicted products. 

For instance, let's consider the previous example for NDCG. Let's assume that among the predicted products for $q_i$, only the products of relevant score greater than equal 0.5 are relevant. That means, only $p^{i}_1$, and $p^{i}_3$ are relevants out of the 5 predicted products. Now $prec@1$ is 1, as the 1st predicted product is relevant.
Similarly, $prec@2$ is 0.5, as out of first two products, 1 is relevant. Then, $rel@j$ states that whether j'th predicted product is relevant. If the j'th predicted term is relevant, $rel@j = 1$, otherwise $rel@j = 0$. Finally, MAP is calculated by the following equation for LTR problem.

\[MAP = \frac{1}{n} \sum_{i=1}^{n} \sum_{j=1}^{m^{(i)}}\frac{prec@j * rel@j}{NRI}\] where $NRI$ is the number of relevant products for the $m^{(i)}$ predicted products. We see that, MAP is calculated taking mean of average precisions for all the queries. For the particular query $q_i$, average precision (AP) = (1*1 + 0.5*0 + 0.6666*1 + 0.5*0 + 0.4*0) / 2 = 0.833
Note that, AP of the predicted products is lower than NDCG score in this particular case.

\subsection{MRR} For LTR problem, the position of the correct products are important. While MAP gives a general idea of the performance of an algorithm, the position of the correct products are ignored. Mean Reciprocal Rank can deal with this issue. The reciprocal rank of a query, $q_i$ is the multiplicative inverse of the rank of the first correct answer. If $rank_i$ demonstrates the rank position of the first relevant product, mean reciprocal rank corresponds to the harmonic mean of the rank positions. MRR is illustrated by the next equation.

\[MRR = \frac{1}{n} \sum_{i=1}^{n} \frac{1}{rank_i}\]

To illustrate with the previous example, for $q_i$, the rank position of the first correct predicted product is 1, and so $rank_1 = 1$. Let's there be $n=3$ queries in the dataset, and for the second query we need to look up top two predictions for the first relevant product. That means the first product is not relevant, and $rank_2 = 2$. Similarly, let's assume $rank_3 = 3$. Then MRR = 1/3 * (1+0.5+0.333) = 0.611

\begin{table}

 \centering\caption{ERR calculation table}

    \begin{tabular}{c|  c|  c | c}\hline
    \bf r & \bf $R_r$ & \bf P(r) & \bf 1/r * P(r)  \\\hline
    1 & 0.34  & 0.34 & 0.34\\
     \hline
    2 & 0.21  & 0.14 & 0.07\\
     \hline
    3  & 0.27 & 0.15 & 0.05\\
     \hline
    4  & 0.15 & 0.07 & 0.035\\
     \hline
    5  & 0.21 & 0.09  & 0.018\\
    \hline
    \end{tabular}
 
\label{table:err-cal}
\end{table}

\subsection{ERR} As MRR works only with the first correct predicted document, Expected Reciprocal Rank (ERR) ~\cite{err2009} is proposed. ERR also diminishes the drawback of DCG such that DCG has an additive nature. Additionally, an assumption of DCG is that a product in a given position has always the same gain and discount independently of the products shown above it. Diminishing these drawbacks, ERR can correlate user click behavior better than DCG, and MRR. To illustrate ERR, at first a probability function is described to measure the satisfaction of a user from the j'th product of $q_i$. Let the relevance of the j'th product $p^{i}_j$ is $rel^{i}_j$. $R_j$ describes the user satisfaction of product $p^{i}_j$ such that the higher the relevance score, the higher the probability of satisfaction.

\[R_j = \frac{2^{rel^{i}_j}-1}{\max(rel^{i})}\]
Here $\max(rel^{i})$ symbolises the maximum relevant score for all the products associated with $q_i$
Then the likelihood of user stops at $r$'th position is calculated based on the $R_j$. Basically, user stops at position $r$, because the previous $r-1$ products do not satisfy him, and the product in position $r$ is satisfactory.
So the likelihood is calculated by the next equation.

\[P\text{(user stops at position r)} = \prod_{j=1}^{r-1} (1-R_j) R_r\]

Finally, ERR is calculated as an expected metric, penalizing the position of a product by the following equation.
\[ERR = \sum_{r=1}^{r} \frac{1}{r} P\text{(user stops at position r)}\] 

To illustrate with an example, let's consider our example where for $q_i$ the relevant scores of the five products are 0.6, 0.4, 0.5, 0.3, and 0.4 respectively.  Table~\ref{table:err-cal} describes the calculation of ERR for these relevance scores. The rightmost column calculates the value of $\frac{1}{r} P\text{(user stops at position r)}$. If we sum over the values of the rightmost column ERR = 0.513 is obtained. Note that, the ERR value is less than any metrics so far we described.

\section{Dataset Availability} \label{dataset} In the field of e-commerce, finding dataset for LTR is very hard. The primary reason is that
such datasets contain confidential or proprietary information, and e-commerce
platforms do not wish to take this risk. To cite an example,\cite{Severyn2015LearningTR,bi2019,brenner2018}worked on LTR with Walmart and Amazon datasets, but the datasets are not published. There are a few datasets for clustering, or recommendation systems, but as the number of features are very less (e.g. 8 features in~\cite{chen2012}), these datasets are unworthy to be worked with LTR problems. To our best knowledge, there exists only a single dataset (Mercateo dataset)\footnote{https://github.com/ecom-research/CRM-LTR}~\cite{anwar2021} for e-commerce LTR. The dataset is consist of query-title pair embedding and relevance score. The queries, and products are collected from search logs of e-commerce platform, and the embedding is performed using Glove~\cite{pennington2014glove}. Raw text for queries, and titles are not published
because the sellers have the proprietary rights on the product title and description. The relevance score is calculated using click rate, and normalization. Alongside the embedding some extra features of the data such as price, delivery time, profit margin etc. The statistics of the dataset is illustrated in left side of the Table~\ref{table:stat}.

The Mercateo dataset has some limitations, as only the Pointwise approaches can be fit with this. The Pairwise and the Listwise approaches are not well fit because as the query, and products are tightly coupled based on embedding, adequate flexibility is missing. So the issue with the scarcity of available datasets persist for e-commerce LTR. That is why in this paper we use an eBay dataset of LTR. For similar reason of data privacy, we can not publish the dataset.

\begin{table}
\caption{Statistics for the Mercateo dataset (left), and the eBay dataset (right)}
\resizebox{0.47\textwidth}{!}{
\begin{tabular}{c|c}
    \hline
    \bf Total \# of Queries &  3060 \\
     \hline
    \# of Queries in [Train/Dev/Test] set  &  [1836/612/612] \\
     \hline
    \# of Products in Dataset set  & 3507965 \\
    \hline
     Avg. \# of Products per Query [Train/Dev/Test] & [1106/1218/1192] \\
    \hline
    \end{tabular}}
\quad
\resizebox{0.47\textwidth}{!}{
\begin{tabular}{c|c}
    \hline
    \bf Total \# of Queries &  44679 \\
     \hline
    \# of Queries in [Train/Dev/Test] set  &  [26807/8936/8936] \\
     \hline
    \# of Products in Dataset set  & 787408 \\
    \hline
     Total \# of Products per Query [Train/Dev/Test] & [20/20/20] \\
    \hline
    \end{tabular}}
\label{table:stat}
\end{table}

The eBay dataset of LTR which we also refer as e-commerce dataset contains three columns, i.e., query, product and the similarity between them. The similarity value can also be described as relevance value which is calculated as a function of click rate and purchase rate. The higher the similarity value, the higher is the product relevant to the corresponding query. Additionally, this value can be used to rank the products. The e-commmerce dataset is consist of 44679 \# of queries, which is much higher than Mercateo dataset. However, the \# of products per query is kept fixed. The ratio of the train, test and dev partition is 6:2:2. Although the number of queries are higher than Mercateo dataset, as the \# of products per query are very less, the total \# of products is around 78 k. The right side of the Table~\ref{table:stat} illustrates the e-commerce dataset. Note that \# of queries per query is 20. 

Finally, the query-product similarity distribution of the e-commerce dataset is shown in Figure~\ref{fig:sim-query}. The similarity value range is between [0.4,1), as the dataset contains only the top 20 similar products. For this histogram we use the same bin size 20 for train, test and dev datasets. It appears that the similarity values follow normal distribution, and around 80\% data fall between range [0.6-0.8]. Additionally, the data distribution among train, test and dev dataset looks similar, that means there is less similarity bias among them.

\begin{figure*}[]%
    \centering
    \subfloat[\centering Train Data Distribution]{{\includegraphics[width=4.5cm, height=3cm]{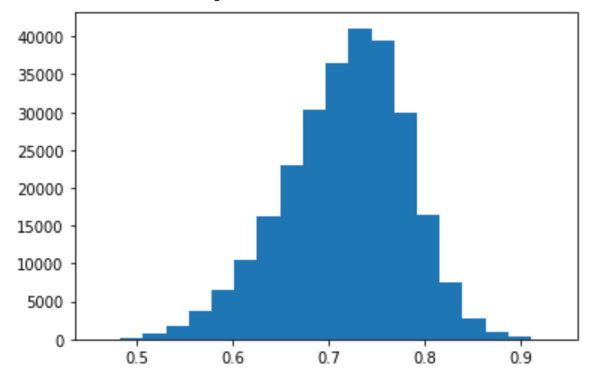} }}%
    \subfloat[\centering Dev Data Distribution]{{\includegraphics[width=4.5cm, height=3cm]{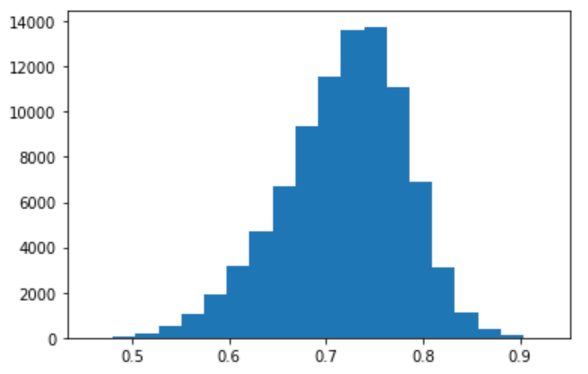} }}%
    \subfloat[\centering Test Data Distribution]{{\includegraphics[width=4.5cm, height=3cm]{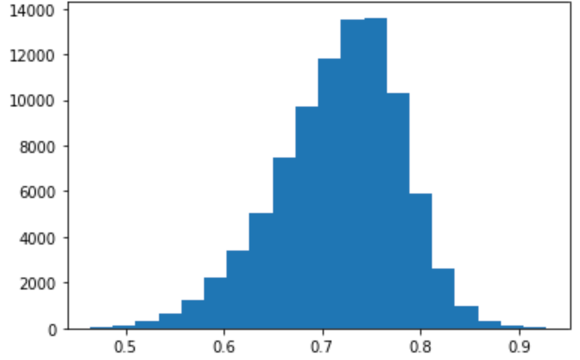} }}%
    \caption{Query-Product Similarity Value Distribution of the Train, Dev, and Test Partitions of the e-commerce Dataset}%
    \label{fig:dd}%
    \hspace{-1.0in}
\end{figure*}

\section{Experiment and Result} In this section we illustrate the performance of all the learning to rank methods which we describe in 
section~\ref{exp}. However, there is not a single research which performs experiment of all the methods we describe in this paper for a single dataset. That's why to compare among the methods is a challenging task. Hence we choose to perform experiments of all the methods in a common e-commerce dataset which we illustrate in the previous section. In the next subsections, we focus the query and product representation first for the experiments, then we illustrate the hyperparameters which we tune for the models, and finally we compare all the methods based on different evaluation metrics.

\subsection{Representation Learning}
Representing an e-commerce query or title is not similar to other domain as they have distinctive characteristics. To perform the experiments with the described methods, each query and product need to be represented in real space such that the similarity between query and product can be captured. In this paper, we only use the query text, and product title text for embedding. While there are many a method~\cite{spherical:Embedding,word2vec-2013,fasttext:Embedding,bert:Embedding} available for text embedding, we use BERT, and Spherical Text Embedding which are most promising capturing the representation of words, and phrases specially for similarity prediction~\cite{ordsim_mdkabir}, and relation understanding~\cite{kabir_Asper,patterncausality}.

\subsubsection{BERT and eBERT}
Bidirectional Encoder Representations from Transformers (BERT) is proposed by a team of researchers from Google~\cite{bert:Embedding} which is not trained on any specific downstream task but instead on a more generic task called Masked Language Modeling. The idea is to uplift huge amounts of unlabeled data to pre-train a model on language modeling. Predicting the next word(s) given a context already requires understanding language to some extent. Next, this pre-trained model can be fine-tuned to solve different kinds of NLP tasks by adding a task specific layer which maps the contextualized token embeddings into the desired output function.

eBERT\footnote{https://www.enterpriseai.news/2021/09/15/heres-how-ebay-is-using-optimization-techniques-to-scale-ai-for-its-recommendation-systems/} is an e-commerce specific version of the BERT model. Along with the Wikipedia corpus, 1 billion latest unique product titles are collected to train the model. The eBERT model can represent the e-commerce terms better than just using only Wikipedia corpus.  In this work, to train the models, we use eBERT to embed both query and product title text.
We keep the embedding dimension of eBERT similar to BERT which is 768.

\subsubsection{Spherical Text Embedding}
Directional similarity is often more effective in tasks, such as, word similarity and document clustering. When textual units
are embedded in the Euclidean space, two textual units may have zero directional distance, yet they are far from each other
by Euclidean distance. This is not desirable when we are trying to predict the query similarity values, which are between
0 and 1, and was computed by directional similarity (Cosine) of two vectors. To overcome this, spherical text embedding 
\cite{spherical:Embedding} has been proposed, which embed the textual units on the surface of a unit d-sphere 
(d is the dimension). To learn embedding on unit d-sphere, an efficient optimization algorithm is proposed with convergence guarantee based on Riemannian optimization. Spherical text embedding shown to be highly effective on various text
embedding tasks, including word similarity and document clustering. For embedding e-commerce query text, we also choose
spherical text embedding as a candidate. To build the corpus of spherical text embedding, besides Wikipedia sentences, we introduce 10 million e-commerce queries from eBay. We choose the dimensionality of the sphere as 100 as
recommended by the original authors. We use each word as a textual unit for embedding task and then apply mean pooling over 
the words of a query to obtain query embedding vectors.

\begin{table*}[!t]
\caption{Comparison with Baseline Methods for all the representations}
\small
\setlength{\tabcolsep}{6pt}
\renewcommand{\arraystretch}{0.7}
        \centering
     
                \begin{tabular}{l|c|c|c|c|c|c}
                    \hline
                    \hline
                    \bf Category & \bf LTR Methods & \bf Embedding & \bf NDCG@10 & \bf NDCG@20 & \bf ERR@10 & \bf ERR@20 
                    \\\hline
                    \hline
                     &  \multirow{2}{*}{LR} & eBERT & 0.627 & 0.623 & 0.513 & 0.511 \\
                     & & Spherical & 0.643 & 0.641 & 0.523 & 0.521
                     \\\cline{2-7}
                     &  \multirow{2}{*}{PR} & eBERT & 0.628 & 0.629 & 0.515 & 0.511 \\
                     & & Spherical & 0.649 & 0.647 & 0.532 & 0.538
                     
                     \\\cline{2-7}
                      Pointwise &  \multirow{2}{*}{SVR} & eBERT & 0.639 & 0.641 & 0.521 & 0.523\\
                     & & Spherical & 0.655 & 0.661 & 0.539 & 0.541
                     \\\cline{2-7}
                      &  \multirow{2}{*}{RF} & eBERT & 0.643 & 0.643 & 0.527 & 0.529\\
                     & & Spherical & 0.657 & 0.656 & 0.547 & 0.545
                     \\\cline{2-7}
                     
                     &  \multirow{2}{*}{\bf GB} & eBERT & 0.682 & 0.691 & 0.553 & 0.557\\
                     
                     & & \bf Spherical & \bf 0.712 & \bf 0.711 & \bf 0.592 & \bf 0.597\\
                     \hline
                     \hline
                     &  \multirow{2}{*}{SVM} & eBERT & 0.641 & 0.642 & 0.523 & 0.525\\
                     & & Spherical & 0.659 & 0.663 & 0.539 & 0.543
                     
                     \\\cline{2-7}
                     
                     &  \multirow{2}{*}{RankNet} & eBERT & 0.657 & 0.655 & 0.541 & 0.539 \\
                     & & Spherical & 0.672 & 0.67 & 0.551 & 0.552
                     \\\cline{2-7}
                      Pairwise &  \multirow{2}{*}{LambdaRANK} & eBERT & 0.671 & 0.673 & 0.56 & 0.567\\
                      & & Spherical & 0.705 & 0.703 & 0.581 & 0.584
                     \\\cline{2-7}
                      &  \multirow{2}{*}{FRank} & eBERT & 0.664 & 0.669 & 0.546 & 0.548\\
                     & &  Spherical &  0.68 &  0.683 &  0.572 &  0.576
                     
                     \\\cline{2-7}
                     &  \multirow{2}{*}{\bf LambdaMART} & eBERT & 0.687 & 0.688 & 0.59 & 0.61\\
                     & & \bf Spherical & \bf 0.715 & \bf 0.714 & \bf 0.64 & \bf 0.643 \\
                     
                     \hline
                     \hline
                     &  \multirow{2}{*}{ListNet} & eBERT & 0.702 & 0.709 & 0.569 & 0.57\\
                     & & Spherical & 0.719 & 0.713 & 0.644 & 0.646
                     \\\cline{2-7}
                      &  \multirow{2}{*}{\bf ListMLE} & eBERT & 0.716 & 0.713 & 0.641 & 0.64\\
                     & & \bf Spherical & \bf 0.723 & \bf 0.725 & \bf 0.651 & \bf 0.652 
                     \\\cline{2-7}
                     \bf Listwise &  \multirow{2}{*}{p-ListMLE} & eBERT & 0.705 & 0. 691 & 0.579 & 0.567\\
                     & & Spherical & 0.711 & 0.702 & 0.609 & 0.594
                     \\\cline{2-7}
                     
                     &  \multirow{2}{*}{DeepQRank} & eBERT & 0.652 & 0.653 & 0.539 & 538\\
                     & & Spherical & 0.669 & 0.663 & 0.549 & 0.55\\
                     \hline
                \end{tabular}
            \label{table:results}
\end{table*}

\subsection{Discussion of Hyper-parameter Tuning} Table~\ref{table:results} shows the performance of the described methods. We only show the best performance of a model. In this section, we want to illustrate the hyperparameter tuning of the models. Note that the hyperparameters of each model is not similar to other models. However, all the pointwise approaches work with replicating the floating point similarity score, the pairwise approaches work with the relative ordering between a pair of products, and the listwise approaches mimic the ranking of a whole list of products. For all the methods, we do not use any default normalization. Additionally, for representation of query text and titles we either use spherical embedding for text or eBERT for embedding the text into vectors. Finally, for all the methods we use mean pooling of the word token embedding from eBERT or spherical text embedding for query or title embedding. Concatenated query and title embedding is the feature for all the methods of this research.

To discuss the hyperparameters of the pointwise approaches we start with LR. There are not many ways to tune the hyperparameters of LR. However, we did not use any kernel fitting over data or $l_{1}$ or $l_{2}$ normalization.
For PR, we tune the degree from 2 to 4, and get best evaluation metrics for degree 2.
For SVR we use ``linear'' kernel and tune parameter $C$ over range [0.5,2.0] with an interval of 0.1. The best result was found for $C=1.1$. For RF regression while there can be quite a few parameters we tune maximum depth of a tree regressor for the values {2,3,4,5} and get the best result for the depth 2. Finally, for GB we tune the maximum depth parameter for the same values like RF. Additionally, we tune the learning rate for the values in the range [0.005, 0.05] with an interval of 0.005. We get the best result for max depth 2 and learning rate 0.015.  

For RankNet, LambdaRANK, and FRank in Table~\ref{table:results} only the loss function changes over methods. The hyperparameters can be model selection, and learning rate. For the learning rate we use ``adam'' optimizer with default learning rate 0.01. We use two hidden layers model for these pairwise methods varying the number of neurons over the values of set {32, 64, 128, 256, 512, 1024}. For RankNet and LambdaRANK number of corresponding neurons 512, 32 provide the best model, while for FRank the number of neurons of the two layers is 256, and 128 for the optimum model.
For RankNet, LambdaRANK, and FRank, the final layer is a dense layer with sigmoid loss to achieve the desired goal.
Next for SVM we tune $C$ over range [0.5,2.0] with an interval of 0.1, and $C=1$ provides the best result. We do not use any hidden layer for SVM, and LambdaMART. For LambdaMART the tunable parameters are number of trees and learning rate. We tune learning rate between 0.01 to 0.2 with an interval of 0.01. Furthermore, number of trees are tuned in the range [2,10] with an interval of 1. When learning rate is 0.08, and number of trees is 3, the best result is achieved.

Likewise Pairwise methods, the first three methods differ only based on loss function; only the MLP parameters are variable. The MLP structure is similar to pairwise approaches as well for an apple to apple comparison. For all the ListNet, ListMLE and p-ListMLE methods we get best result when number of neurons in the first and second hidden layers are 256, and 128 respectively. Note that, the output layer of the first three methods depend on how many products to be ranked by a model per query. All the losses are calculated based on the output value of a product to its corresponding query. For all the three methods, to calculate similarity value we use both dense layer and cosine similarity. The dense layer value is a better choice for our dataset in terms of better performance because of the flexibility of a dense layer over cosine similarity. Finally, for DeepQRank we use the default parameters of the original project.

\subsection{Result} In this subsection we provide an extensive evaluation of all the methods. To evaluate we use NDCG@K, and ERR@K for K=10, and K=20. Note that the other evaluation metrics are not well suited for e-commerce LTR methods. Among all the stated methods, pointwise approaches are the best suit if the number of candidate products to be ranked are higher. In contrast, the pairwise approach, and listwise approaches are better suited when there are suggestions of products to be ranked. For every method Spherical text embedding is a better representation than eBERT, which is better suited to similarity and ranking based tasks~\cite{ordsim_mdkabir}. For all the methods, there seems to be a positive correlation between NDCG and ERR. One possible reason can be both the models are good fit for e-commerce LTR evaluation.

Among the pointwise approaches, both LR and PR performs almost similarly for the e-commerce dataset, and the LTR performance based on both NDCG and ERR are lowest for them. SVR improves the result, but the magnitude of improvement is not very high. One possible reason can be all these models are shallow, and the features from embedding space are not well synchronised for ranking. RF improves both NDCG and ERR scores, while GB is the best among pointwise approaches. NDCG@10 for GB with Spherical embedding is 0.712 which is 11\% improvement over LR with the same embedding method. Similarly, with respect to ERR@10 the improvement is 13\%. It is evident from the result that GB outperforms all other pointwise approaches. Note that GB is a tree and boosting based regression method. Output of the shallow tree contributes the predicted score, and possibly not all the features are equally important, and GB better capture that compared to RF. Finally, for K=10, and K=20 the performance of the same pointwise approach is to some extent similar. As the pointwise approaches deal with actual similarity value, changing K does not effect the performance much.

Next the pairwise approaches overall perform better than the pointwise approaches. SVM performs similar to SVR of the pointwise approach which is expected because the feature space remains same and both model are aiming to minimize similar type of loss functions, although not exactly similar. While RankNet, FRank, and LambdaRANK work with the same model and different loss functions, LambdaRANK performs better. The results denote that both FRank and LambdaRANK provide better loss function than RankNet. The reason can be LambdaRANK provides better loss function among these three. However the best pairwise method is LambdaMART. LambdaMART is a boosted tree version of LambdaRANK, and the output of the weak learners are gathered nicely for final prediction. The model is optimized using gradient boosting. These strengths may lead LambdaMART best pairwise method for the e-commerce dataset.

Finally, among the listwise approaches we work with, DeepQRank is not a good fit with our dataset. Note that DeepQRank is a reinfocement based Markov model, and the reward function is maximized based on maximizing the reward function. But actually a shallow neural network is trained to maximize the re-reinforcement learning. For the e-commerce dataset, the loss function is similar to pointwise approaches as the complete list of products of a query is not considered at a time for training. All other listwise approaches performs comparatively better as the loss function is designed for the whole permutation. Moreover, totally irrelevant products are ignored in listwise approach, which allows the model to concentrate on the actual relevant products. While there is a slight difference in the performance of ListNet, ListMLE, and p-ListMLE methods, we get best score for ListMLE. All of these methods are similar except the loss function. All of these three methods work well in other datasets with some pros and cons. The ListNet is a good fit for the datasets where number of required products to be ranked is smaller than both ListMLE and p-ListMLE. On the other hand, p-ListMLE is a good fit when there is emphasize on the top products. In our case we have k=10, and 20 which are not very large, and that might be a cause for ListMLE to be performed better.

To conclude, the performance shown in the Table \ref{table:results} advocates the best methods of learning to rank as Gradient Boosting, LambdaMART, and ListMLE. While for the eBay dataset the best among these three is ListMLE, this might change in other datasets. For example XGBoost provides the best rest result for CIKM Cup 2016~\cite{Palotti2016LearningTR}.

\section{Conclusion and Future Work}
While there are quite a few number of research works available in web domain, recommender system and other IR domains, tendency of publishing models for e-commerce domain remains confidential. The LTR methods face specific challenges when implemented in the e-commerce domain. Additionally, although e-commerce LTR is very important as understanding users' intent is important to increase the purchase rate for a client, starting with LTR methods are still a puzzle. In this research, we want to provide a general idea for starting road-map. We illustrate the methods in details so that each method is easily understandable and implementable. Additionally, we perform experiments with the illustrated methods so that an idea can be generated to select from the described methods.
Our another goal is to share the codes as soon as the paper is published so that LTR in e-commerce domain can easily be performed. In future, we want to work on more sophisticated deep learning methods, which can improve the performance substantially.

\bibliographystyle{elsarticle-num}
\bibliography{main}

\end{document}